%% file: main.tex
\documentclass[aps,prd,twocolumn,superscriptaddress,showpacs,amsmath,amssymb,floatfix,preprintnumbers,nofootinbib]{revtex4-2}
\usepackage[dvipdfmx]{graphicx}
\usepackage{graphicx}% Include figure files
\usepackage{dcolumn}% Align table columns on decimal point
\usepackage{bm}% bold math
\usepackage{soul}
\usepackage{multirow}
\usepackage{color}
\usepackage{url}
\usepackage{xcolor}
\usepackage[italic]{hepnames}
\usepackage{tabularx}
\usepackage{array,booktabs}
\usepackage{caption}
\usepackage{subcaption}
\usepackage{adjustbox}
\usepackage{placeins}
\usepackage{lineno}
\usepackage{orcidlink}
\definecolor{BlueViolet}{rgb}{0.2, 0.00, 0.7}
\definecolor{Blue}{rgb}{0.15, 0.00, 0.9}
\usepackage[scientific-notation=true]{siunitx}
\input belle2sym.tex

\begin{document}

\preprint{\vbox{ \hbox{   }
                        \hbox{Belle Preprint 2023-20}
                        \hbox{KEK Preprint 2023-38}
}}

\title{Angular analysis of $B \to K^* e^+ e^-$ in the low-$q^2$ region with new electron identification at Belle}

\noaffiliation

\input{pub665-orcid.tex}

\begin{abstract}
We perform an angular analysis of the $B\to K^* e^+ e^-$ decay for the dielectron mass squared, $q^2$, range of $0.0008$ to $1.1200 ~\text{GeV}^2 /c^4$ using the full Belle data set in the $K^{*0} \to K^+ \pi^-$ and $K^{*+} \to K_S^0 \pi^+$ channels, incorporating new methods of electron identification to improve the statistical power of the data set. This analysis is sensitive to contributions from right-handed currents from physics beyond the Standard Model by constraining the Wilson coefficients $\mathcal{C}_7^{(\prime)}$. We perform a fit to the $B\to K^* e^+ e^-$ differential decay rate and measure the imaginary component of the transversality amplitude to be $A_T^{\rm Im} = -1.27 \pm 0.52 \pm 0.12$, and the $K^*$ transverse asymmetry to be $A_T^{(2)} = 0.52 \pm 0.53 \pm 0.11$, with $F_L$ and $A_T^{\rm Re}$ fixed to the Standard Model values. The resulting constraints on the value of $\mathcal{C}_7^{\prime}$ are consistent with the Standard Model within a $2\sigma$ confidence interval.
\end{abstract}

{
\let\clearpage\relax
\maketitle
}

\tighten

{\renewcommand{\thefootnote}{\fnsymbol{footnote}}}
\setcounter{footnote}{0}

\section{Introduction}
The $\btosll$ decay is a flavor-changing neutral current mediated by loop diagrams and is therefore suppressed in the Standard Model (SM). Studies of this decay are a sensitive probe for contributions from physics beyond the SM, where angular analyses measuring parameters with reduced theoretical uncertainties, including $\pfp$, have in the past found disagreements with SM predictions~\cite{LHCb:2020lmf,Belle:2016fev}. Effective field theory is used to characterize the Hamiltonian and to describe the different processes that contribute to the decay~\cite{Altmannshofer:2008dz}. The contributions from different operators across the range of the dilepton invariant mass squared, $q^2$, are determined by their respective Wilson coefficients. 
In the $q^2$ region where the $\pfp$ tension is found, the $\mathcal{C}_9^{(\prime)}$ Wilson coefficients are dominant, with some contributions from $\mathcal{C}_7^{(\prime)}$ and from charm loops that are not yet fully understood~\cite{Khodjamirian:2010vf}.

Studies focusing on the very low-$q^2$ region, $q^2 \lesssim 1.0 \gevccsq$, are accessible for the dielectron mode and can isolate the $\mathcal{C}_7^{(\prime)}$ contribution, which measures the polarization of photons from $b \to s \gamma$ decays and is a probe for non-SM right-handed currents~\cite{Becirevic:2012dx}. Angular analyses of $\btosll$ can be used to test other new physics scenarios, such as those detailed in Refs.~\cite{Kou:2013gna, EBERL2021PRD, Blanke:2012tv,Malm:2015oda}. The $B^0 \to K^{*0}(\to K^+ \pi^-) e^+ e^-$ decay at low-$q^2$ can be used to measure the ratio of the right- and left-handed Wilson coefficients, $\csev^\prime /\csev$, which constrains non-SM contributions. Such angular analyses have been performed by LHCb for $0.002 < q^2 < 1.120 \gevccsq$~\cite{LHCb:2015ycz} and $0.0008 < q^2 < 0.257 \gevccsq$~\cite{LHCb:2020dof}, where results were consistent with the SM. 
The following study is an independent angular analysis of the very low $q^2$ region in Belle, and the first of its kind at an $e^+ e^-$ $B$-factory.
% A recent study of lepton flavor universality in $\BtoKstee$ decays in different $q^2$ ranges from LHCb~\cite{LHCb:2022qnv} used tighter electron identification requirements than previous analyses~\cite{LHCb:2017avl,LHCb:2021trn} and obtained measurements more consistent with the SM compared to those previous analyses. This highlights the importance of robust systematic uncertainty studies in rare decays and the need for a low-$q^2$ analysis in Belle as an independent measurement of other angular analyses involving electrons. 
% To gain a better understanding of $\btosll$ decays, analyses are being re-performed with improved systematic uncertainty studies, for example the test of lepton flavor universality at LHCb in the low $q^2$ region. This study obtained measurements closer to SM expectations~\cite{LHCb:2022qnv} than previous analyses~\cite{LHCb:2017avl,LHCb:2021trn}. It is therefore important to perform for the first time an independent angular analysis of the very low $q^2$ region with Belle, with a robust evaluation of systematic uncertainties.
To offset the smaller available data set in Belle, this study includes the implementation of machine learning techniques to improve the performance of electron identification. 

\subsection{Differential decay rate}
The $\BtoKstee$ differential decay rate is expressed in terms of $q^2$ and three angular observables, $\theta_{\ell}$, $\theta_{K}$, and $\phi$~\cite{LHCb:2013zuf}. The angle $\thetal$ is defined as the angle between the $e^+$ ($e^-$) candidate direction and the direction opposite to the $B$ ($\Bbar$), in the dielectron rest frame. The angle $\thetaK$ is defined similarly between the kaon direction and the opposite direction of the $B$ in the $\Kstar$ rest frame. Finally, $\phi$ is the angle between the plane containing the dielectron candidates and the plane containing the kaon and pion candidates in the $B$ rest frame. In the low-$q^2$ region, the full differential decay rate~\cite{Altmannshofer:2008dz} can be simplified by folding the $\phi$ distribution, which is done by adding $\pi$ to values of $\phi$ below $0$. The $K\pi$ S-wave contribution can be neglected in this $q^2$ region due to the polarisation of the hadronic system~\cite{LHCb:2016ykl}, and electrons can be taken to be massless~\cite{LHCb:2020dof}. The differential decay rate, assuming there are no scalar or tensor contributions, is averaged over $CP$-conjugate modes throughout this paper and is given by
\begin{equation}
\label{eq:diff_reduced}
\begin{aligned}
    &\frac{1}{d(\Gamma + \bar{\Gamma})/dq^2} \frac{d^4 (\Gamma + \bar{\Gamma})}{dq^2 d \cos{\thetal} d \cos{\theta_K} d \phi} = \\
    &\frac{9}{16 \pi}  \Big(\frac{3}{4}(1 - F_L) \sin^2\thetaK + F_L\cos^2 \thetaK \\
    &+ \big( \frac{1}{4}(1-F_L)\sin^2 \thetaK - F_L \cos^2 \thetaK \big) \cos 2\thetal \\
    &+ \frac{1}{2}(1-F_L)\ATt \sin^2 \thetaK \sin^2 \thetal \cos 2\phi \\
    &+ (1-F_L)\ATR \sin^2 \thetaK \cos \thetal \\
    &+ \frac{1}{2}(1-F_L) \ATI \sin^2 \thetaK \sin^2 \thetal \sin 2 \phi \Big),
\end{aligned}
\end{equation}
with four free parameters as functions of $q^2$: the longitudinal polarization of the $\Kstar$, $F_L$, the transverse asymmetry of the $\Kstar$, $\ATt$, and the real and imaginary components of the transversality amplitudes, $\ATR$ and $\ATI$, respectively.
The parameters $\ATI$ and $\ATt$ are related to $\csev^{(\prime)}$ at $q^2 = 0$ through the following equations~\cite{Becirevic:2011bp}:
\begin{align}
    \ATI(q^2=0) &= \frac{2 \text{Im}(\csev^{\text{eff}}\csev^{\prime \text{eff} *})}{|\csev^{\text{eff}}|^2+|\csev^{\prime \text{eff}}|^2},\\
    \ATt(q^2=0) &= \frac{2 \text{Re}(\csev^{\text{eff}}\csev^{\prime \text{eff} *})}{|\csev^{\text{eff}}|^2+|\csev^{\prime \text{eff}}|^2},
\end{align}
where $\csev^{(\prime)\rm eff}$ is proportional to $\csev^{(\prime)}$. In the SM, $\csev^{\prime}$ is helicity-suppressed by a factor of the ratio of the strange and bottom quark masses, $\csev^{\prime} = (m_s/m_b) \csev$~\cite{Altmannshofer:2008dz}, hence these parameters are expected to be near zero at $q^2 = 0$. Therefore, their measurement is sensitive to new physics scenarios. The other parameters, while not sensitive to new physics themselves, can be used to discern different beyond-SM scenarios.

\subsection{The Belle detector and data sample}
This study uses the full $\Upsilon(4S)$ data sample containing $(772 \pm 11) \times 10^6$ \BB meson pairs recorded with the Belle detector~\cite{Belle:2000cnh,Belle:2012iwr} at the KEKB asymmetric-energy $e^+e^-$ collider~\cite{Kurokawa:2001nw,Abe:2013kxa}.

The Belle detector is a large-solid-angle magnetic spectrometer that consists of a silicon vertex detector (SVD), a 50-layer central drift chamber (CDC), an array of aerogel threshold Cherenkov counters (ACC),  % <- \v{C}erenkov 2007.08
a barrel-like arrangement of time-of-flight scintillation counters (TOF), and an electromagnetic calorimeter comprised of CsI(Tl) crystals (ECL) located inside  a super-conducting solenoid coil that provides a 1.5~T magnetic field.  An iron flux-return located outside of the coil is instrumented to detect $K_L^0$ mesons and to identify muons (KLM).  The detector is described in detail elsewhere~\cite{Belle:2000cnh}. The coordinate system is defined such that the positive $z$-axis aligns with the direction of the electron beam and is centered on the interaction point (IP).

Monte Carlo (MC) simulation studies are used to determine the analysis techniques. The MC samples include on-resonance $\FourS \to \BB$ events and continuum $\epen \to \qqbar$ events with $\q \in \{u,d,s,c\}$, which are generated using the EVTGEN~\cite{Lange:2001uf}, PYTHIA~\cite{Sjostrand:2014zea}, and PHOTOS~\cite{Barberio:1990ms} packages with interference effects due to final-state radiation being switched on. The B2BII package~\cite{Gelb:2018agf} is used to convert reconstructed events into a format compatible with the Belle II analysis software framework~\cite{Kuhr:2018lps}. Samples of $\BtoKstee$ signal events are generated using the EVTGEN generators BTOSLLBALL~\cite{Ali:1999mm}, with form factors from Ref.~\cite{Ball:2004rg}, and BTOSLLNP~\cite{Sibidanov:2022gvb}, to study detector effects and understand any possible model dependence in the analysis procedure.

\section{Electron identification}
\label{sec:paper_LID}
Electron identification in Belle is typically performed using a likelihood ratio (LHR), $\mathcal{L}_e/(\mathcal{L}_e + \mathcal{L}_\pi)$, where $\mathcal{L}_i$ is the likelihood of a charged particle hypothesis, $i$, determined using a combination of detector outputs~\cite{eIDBelle}.
We have implemented a new method for electron identification in Belle using machine learning algorithms. A boosted decision tree (BDT) algorithm~\cite{Keck2017} is used to classify an electron signal against all other long-lived charged particle hypotheses, exploiting the properties assigned to candidates by the detectors. BDTs are trained using $500$ trees, a maximum depth of $3$ and a shrinkage rate of $0.1$, in different bins of momentum, angle and charge. For electron identification the BDT input variables are as follows: the binary LHR for hadron identification in the inner detectors~\cite{NAKANO2002402} for all combinations of long-lived charged particles, the ratio of ECL cluster energy and measured momentum, and the ratio of the energy in the $3\times 3$ and $5 \times 5$ ECL crystal grids around the center of an electromagnetic shower. Also included are the ECL cluster energy, the number of crystal hits in an ECL cluster and the lateral shower shape, as defined in Ref.~\cite{DRESCHER1985464}. These variables are assigned to long-lived charged particle candidates that have ECL clusters associated with hits in the CDC.

\subsection{Tag and probe procedure}
The performance and agreement between data and MC for the BDT classifer is verified using a tag and probe method with $J/\psi \to \epen$ events for efficiency measurements and $\KS \to \pipi$ events to determine $\pi$-$e$ mis-identification rates in each phase-space bin. The $\jpsi$ and $\KS$ candidates are taken inclusively from all $\BB$ and continuum events.

$\jpsi$ candidates are reconstructed in a window around the known $\jpsi$ invariant mass~\cite{cite-PDG} from two electron candidate tracks that originate from the IP, where the distance of the tracks from the IP in the $z$-direction must be $|dz| < 5 \cm$, and their radius in the $r$-$\phi$ plane is $|dr| < 2 \cm$. The momentum is required to be $p_{\rm lab} > 0.1 \gevc$ and there must be a match between a track in the CDC and a cluster in the ECL. A match is determined by extrapolating a charged track into the ECL and checking if any crystals it passes through are associated to an ECL cluster. A correction for bremsstrahlung energy loss is applied by adding the four-momenta of photon candidates with $E_{\gamma}<1$ GeV within an angular cone of $6^\circ$ around the electron candidate's track direction from the IP. Low multiplicity $\epen \to (\epen)\ell^+ \ell^-$, and continuum events are reduced through selection criteria based on the event topology. The tagging criterion is a requirement that one $e^\pm$ candidate in the event, the tag, has an $e$ vs. $\pi$ LHR~\cite{eIDBelle}, above $0.95$.

$\KS$ candidates are reconstructed in a window around their known invariant mass using the same track, momentum, and topology requirements as $\jpsi$ candidates. The cosine of the angle between the $\KS$ momentum vector and the decay vertex position vector, $\mathrm{cos}(\theta(\vec{p}_{\KS},\vec{V}_{\KS}))$, is required to be above $0.998$. No pion-tagging LHR criterion is required.

The number of $\jpsi$ and $\KS$ candidates is determined with a binned maximum log-likelihood fit to the dielectron or dipion invariant mass distribution, respectively. For $\jpsiee$, the signal probability distribution function (PDF) is modeled by a Gaussian function added to a bifurcated Gaussian and a Crystal Ball~\cite{cb_function} function, while for $\kspipi$, the signal component is modeled by a sum of three Gaussian functions. In both cases, the background is modeled by a second-order Chebychev polynomial. The shapes of these PDFs are defined based on MC, where the signal and background yields, the means, and a scale factor for the PDF widths remain floating in fits to data with all other parameters fixed.

The efficiency, $\varepsilon$, of applying a selection criterion on the probe is defined as follows:
\begin{equation}
    \varepsilon = \frac{N^{\mathrm{sig}}_{\mathrm{pass}}}{N^{\mathrm{sig}}_{\mathrm{pass}} + N^{\mathrm{sig}}_{\mathrm{fail}}},
    \label{eqn:Jpsi_Ks_eff}
\end{equation}
where $N^{\mathrm{sig}}_{\mathrm{pass}}$ is the number of signal candidates that pass the criterion, and $N^{\mathrm{sig}}_{\mathrm{fail}}$ is the number that do not. A simultaneous fit is performed over the mutually exclusive pass and fail data sets to avoid double counting the statistical uncertainty, where signal shape parameters are common for the two samples while those for the backgrounds are fit independently.

The statistical and systematic uncertainties are determined through the generation of pseudo-experiments.
The statistical uncertainty is calculated in each phase-space bin and is determined by the following procedure. The population in each invariant mass bin is re-sampled according to a Poissonian probability and the invariant mass fit is performed on each new pseudo-data set to obtain a set of efficiencies. The width of the central $68\%$ of a Gaussian PDF fit to this distribution of efficiencies is taken as the statistical uncertainty.
For the systematic uncertainty, which is only assigned to data, an analogous pseudo-experiment method is used, where the fixed signal PDF parameters are varied according to their uncertainty. 
The efficiency ratios between data and MC in different bins, and the corresponding statistical and systematic uncertainties, are then calculated using these pseudo-data sets, and are used for corrections in the $\BtoKstee$ study. An additional study of track isolation effects was performed, finding that although performance degrades when tracks are in close proximity to each other, it is well described by simulation and no additional systematic uncertainty needs to be assigned.

\subsection{Performance comparison}
\begin{figure}
    \centering
    \includegraphics[width=0.98\linewidth]{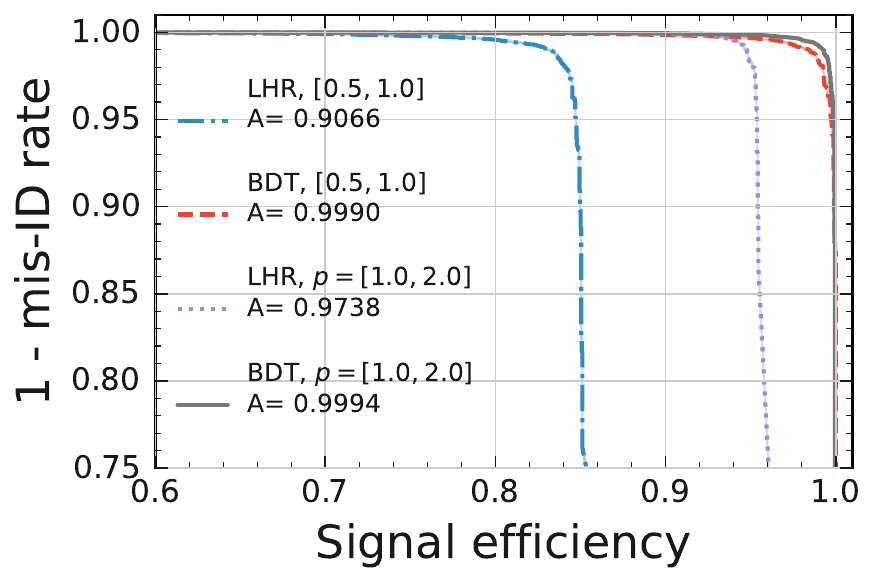}
    \caption{ROC curves comparing the simulated electron identification performance of the BDT and LHR in the ECL barrel region are shown for two different momentum regions, in $\gevc$. The area under the curve, $A$, is also provided for comparison.}
    \label{fig:ROC_bothp}
\end{figure}

A comparison between the performance of the LHR and BDT methods is shown using receiver operating characteristic (ROC) curves for simulated electrons in Fig.~\ref{fig:ROC_bothp}, with the mis-identification, or mis-ID, rate plotted against the signal efficiency. The BDT has a larger area under the curve and therefore outperforms the standard LHR. Similar performance is found for positrons.

For a comparison in data, probe selection criteria thresholds are chosen for each phase-space bin such that the simulated electron signal efficiency is $95\%$, which is in agreement between fits to $\jpsiee$ events for data and MC. The $\pi$-$e$ mis-identification rates in $\kspipi$ are then measured in data using the same criteria. The results for data, integrated over momentum, are presented in Table~\ref{tab:paper_LID}, where a significant reduction in the mis-identification rate is found for the BDT in all angular regions except for the backwards endcap of the ECL, $2.23 < \theta < 2.71 \rad$. 

\begin{table*}[htb]
    \centering
    \begin{tabular}{ccrr}
Charge$~~$ & $~~\theta$ region and range (rad.) & ~~BDT mis-ID in data $\times 10^{3}$ & ~~LHR mis-ID in data $\times 10^{3}$ \\
\hline
   $+$ &         Forward $(0.22,0.56)$ &           $0.80~\pm {0.05}~\pm {0.01}$ &          $5.72~\pm {0.11}~\pm {0.04}$ \\
   $-$ &         Forward $(0.22,0.56)$ &          $0.52~\pm {0.04}~\pm {0.01}$ &           $4.81~\pm {0.12}~\pm {0.03}$ \\
   $+$ &         Barrel $(0.56,2.23)$ &          $0.53~\pm {0.02}~\pm {0.01}$ &          $3.44~\pm {0.03}~\pm {0.01}$ \\
   $-$ &         Barrel $(0.56,2.23)$ &          $0.31~\pm {0.02}~\pm {0.01}$ &          $3.24~\pm {0.03}~\pm {0.01}$ \\
   $+$ &         Backward $(2.23,2.71)$ &         $26.05~\pm {0.28}~\pm {0.35}$ &          $25.30~\pm {0.27}~\pm {0.33}$ \\
   $-$ &         Backward $(2.23,2.71)$ &         $16.73~\pm {0.23}~\pm {0.14}$ &         $18.42~\pm {0.24}~\pm {0.23}$ \\
   $+$ &         Integrated $(0.22,2.71)$ &          $0.78~\pm {0.02}~\pm {0.01}$ &          $4.45~\pm {0.03}~\pm {0.01}$ \\
   $-$ &         Integrated $(0.22,2.71)$ &          $0.54~\pm {0.01}~\pm {0.01}$ &          $4.07~\pm {0.03}~\pm {0.01}$ \\
\end{tabular}
    \caption{The mis-identification rates in data are compared between the BDT and LHR methods. The first errors are statistical and the second are systematic. The results use pions from $\KS \to \pipi$ candidates in data integrated over the full momentum space, measured using a probe selection threshold for the corresponding identification method that results in a $95\%$ signal MC efficiency in $\jpsiee$ candidates.}
    \label{tab:paper_LID}
\end{table*}

It should be noted that this performance is dependent on the presence of a match between a particle's track in the CDC and a shower in the ECL, and by requiring this match, the signal efficiency when using the BDT is reduced. Low-momentum electrons with $p < 1.0 \gevc$ make up $20\%$ of the kinematic phase-space of the $\BtoKstee$ decay, where this track-cluster matching has a reduced efficiency. Therefore for optimal electron identification performance, a combination of the BDT and LHR is used in the following analysis, using the BDT when there is a track and cluster match, and using LHR otherwise. 

\section{$\BtoKstee$ selection and reconstruction}
We reconstruct $\BtoKstee$ events in two channels: the neutral mode with $\Kstarz \to K^+ \pi^-$, and the charged mode with $\Kstarp \to \KS \pi^+$. Pairs of electron and positron candidates are combined with pion and kaon candidates to form $B$ meson candidates. These $e^\pm$ candidates are selected using a combination of the new electron identification BDT discussed earlier ($\text{BDT}_{e}$) and the standard LHR. The thresholds for each charge and identification method, inclusive of momentum and angle, are determined simultaneously from MC by maximising a figure of merit (FOM), given by
\begin{equation}
\label{eq:fom}
    \text{FOM} = \frac{N_S}{\sqrt{N_S+N_B}},
\end{equation}
where $N_S$ is the number of $e^+e^-$ candidate pairs with correctly assigned particle hypotheses and $N_B$ is the number of candidate pairs with any incorrectly assigned hypothesis, after applying the $\text{BDT}_{e}$ and LHR criteria. The thresholds correspond to a signal efficiency of $94\%$ in simulated $\jpsiee$ candidates and the data/MC correction factor is measured to be $1.00 \pm 0.02 \pm 0.01$, where the first uncertainty is statistical and the second is systematic. Therefore the results in Table~\ref{tab:paper_LID} are a good representation of the lepton mis-identification rates in this analysis.

Charged pion candidates are required to have a binary $\pi/K$ LHR~\cite{NAKANO2002402}, $\mathcal{L}_{\pi/K}$, above $0.6$ which retains $95\%$ of true pions and removes $55\%$ of the incorrectly assigned candidates, while charged kaon candidates have $\mathcal{L}_{\pi/K} < 0.9$, which retains $95\%$ of true kaons and removes $80\%$ of the incorrectly assigned candidates. These thresholds are chosen by maximizing the FOM from Eq.~\ref{eq:fom}. Both charged hadrons are required to originate from near the IP, where the distance from the IP in the $z$-direction must be $|dz| < 5 \cm$, while the requirement of their radius in the $r$-$\phi$ plane is $|dr| < 2 \cm$.
$\KS$ candidates are selected from a pair of oppositely charged tracks that form a detached vertex, applying selection criteria for their invariant mass and reconstructed vertex, depending on their assigned momentum~\cite{Fang:2003md}.

The $K \pi$ invariant mass is required to be near the mass of the $\Kstar(892)$ meson, $0.7 < M_{K \pi} < 1.0 \gevcc$. The upper threshold for the dielectron invariant mass is chosen to be $q^2 < 1.12 \gevccsq$ to match the 2015 LHCb study~\cite{LHCb:2015ycz}, as this maximizes the available data sample while remaining near the upper $q^2$ threshold region in which the simplifications of the differential decay rate that give Eq.~\ref{eq:diff_reduced} are valid~\cite{Becirevic:2012dx}.

The beam-energy constrained mass is required to be $M_{{\rm bc}}= \sqrt{E^{*2}_{\rm{beam}}/c^4 - p^{*2}_{B}/c^2 } > 5.23 \gevcc$, where $p^{*}_{B}$ is the momentum of the \B meson and $E^{*}_{\rm {beam}}$ is the beam energy, both in the center-of-mass frame (denoted by the symbol $^*$). The energy-difference variable, $\Delta E = E^{*}_{B} - E^{*}_{\rm {beam}}$, where $E^{*}_{B}$ is the energy of the \B meson, is required to be $|\Delta E| < 0.3 \gev$. In each event, only the candidate with the lowest value of $|\Delta E|$ is retained. 

\subsection{Continuum and signal selection BDTs}
Event shape variables including event kinematics, the ratios of Fox-Wolfram moments~\cite{Wolfram}, and spherical harmonic moments of the momenta of particles are used to train a BDT to classify signal events originating from $\FourS$ decays against continuum background, $\text{BDT}_{\qqbar}$.

\begin{figure*}
    \centering
    \includegraphics[width=0.48\linewidth]{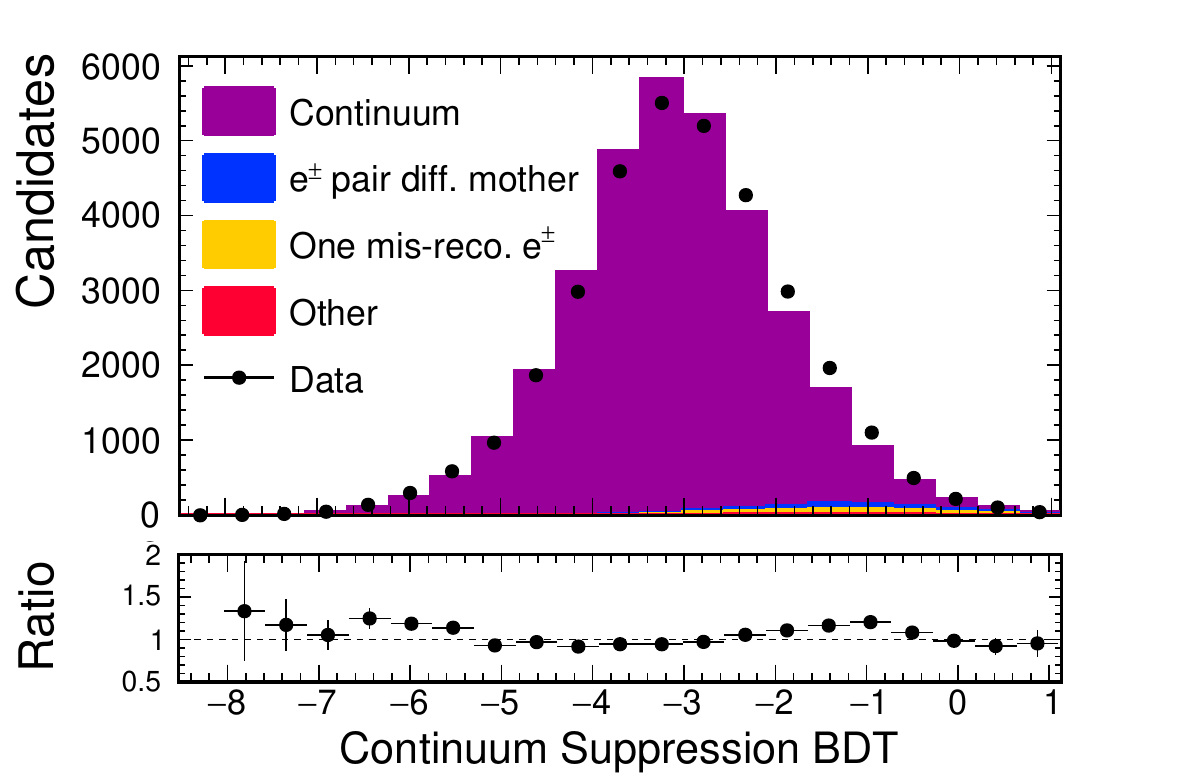}
    \includegraphics[width=0.48\linewidth]{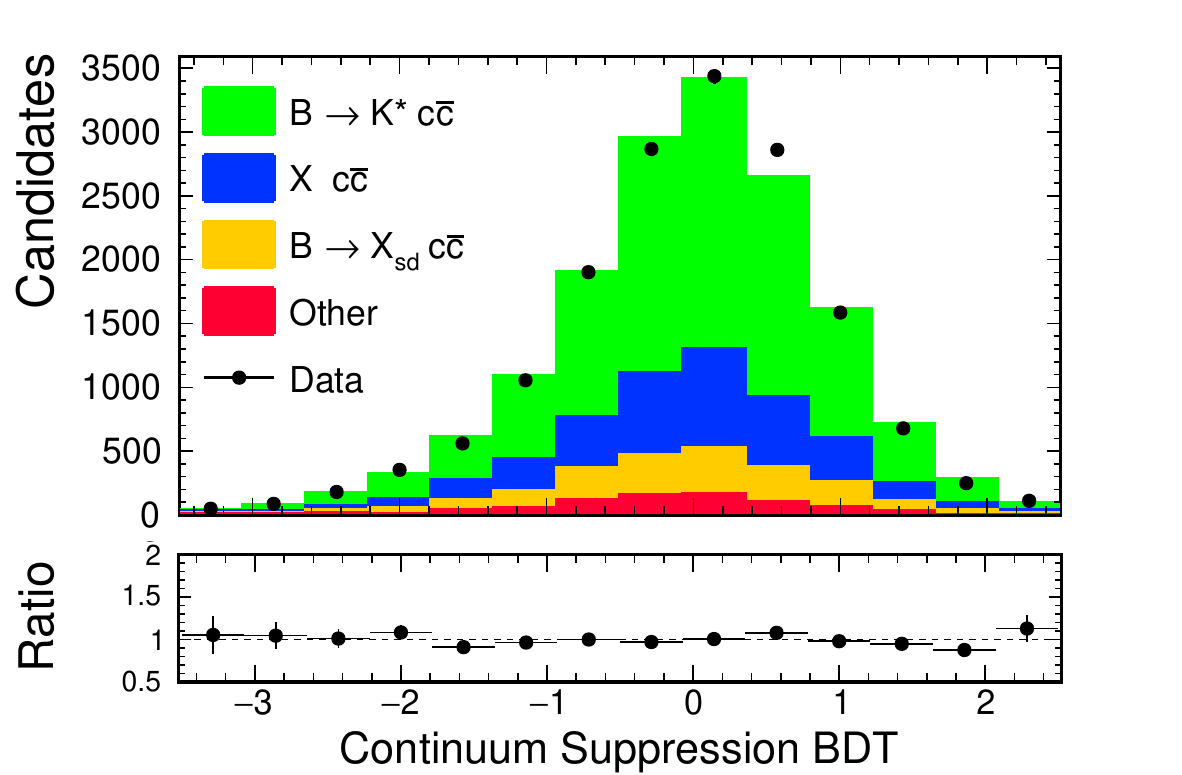}\\
    \includegraphics[width=0.48\linewidth]{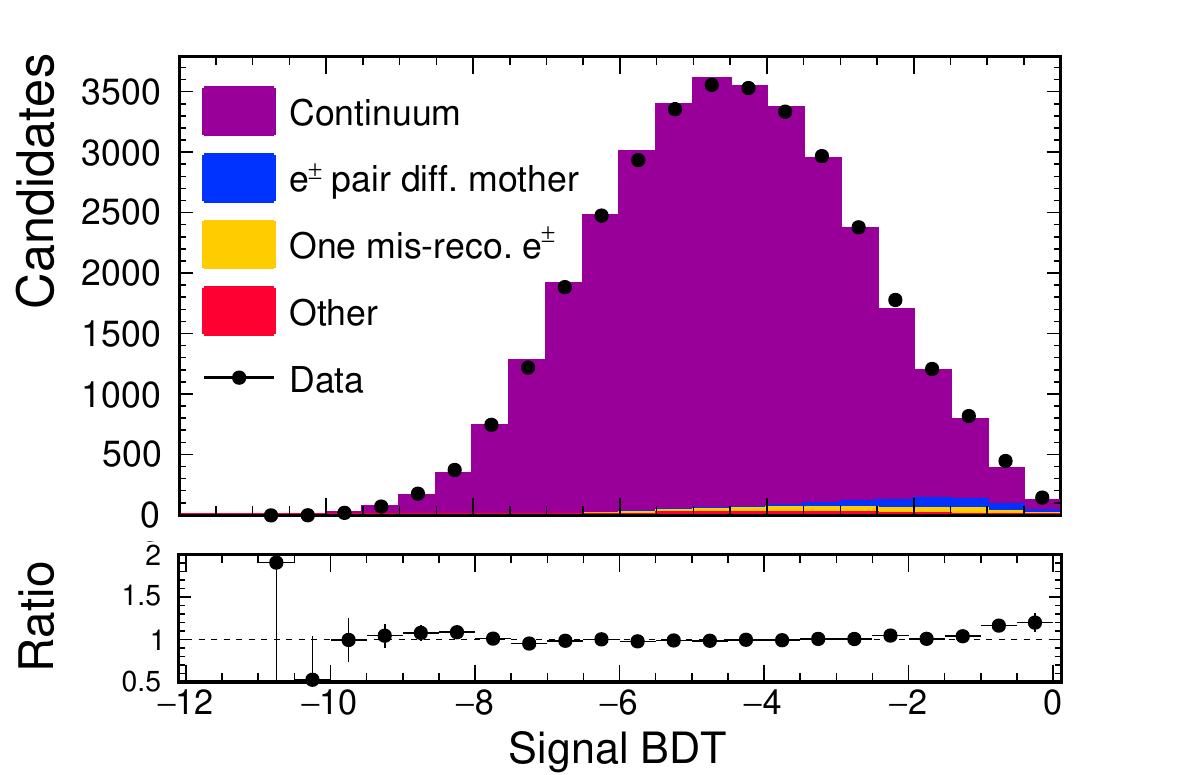}
    \includegraphics[width=0.48\linewidth]{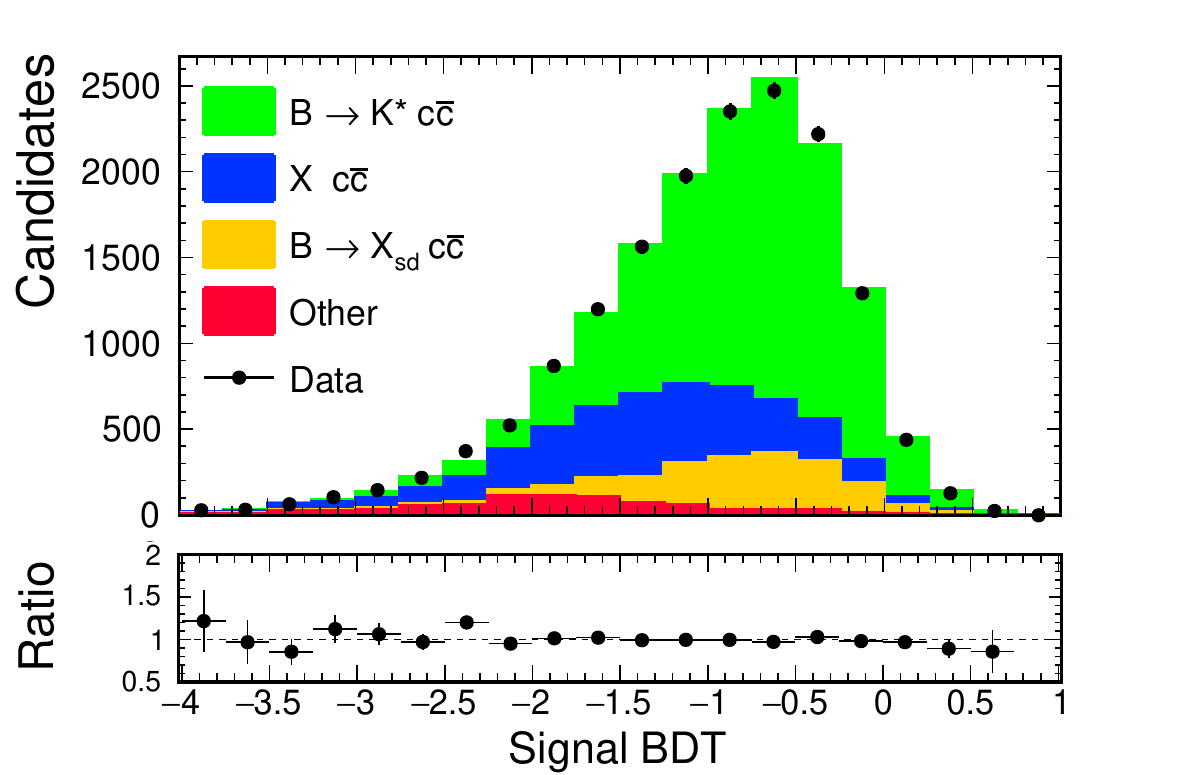}
    \caption{The outputs of $\text{BDT}_{\qqbar}$ (top) and $\text{BDT}_{\text{S}}$ (bottom) in control modes are shown for the $\mbc < 5.27 \gevcc$ sideband (left) and $\B \to \Kstar c\bar{c}$ events (right), comparing the full Belle data set and simulations, scaled to the recorded data luminosity. The particle $X_{sd}$ represents the $S$-wave $K \pi$ contribution, $X \ccbar$ is any mis-reconstructed $\Kstar$ with a correctly reconstructed dielectron pair from a $\ccbar$ resonance, $e^\pm$ pairs with different mothers are any event where the $e^\pm$ candidates were reconstructed correctly but originate from different decays, and the one mis-reconstructed $e^\pm$ classification is applied to any event with a single incorrectly assigned $e^\pm$ candidate. The ratios shown are for the number of candidates in data divided by the number of candidates in the scaled MC.}
    \label{fig:paper_BDTs}
\end{figure*}

The $\text{BDT}_{\qqbar}$ output is then used as one of the inputs to a BDT used to classify signal events against all other background, $\text{BDT}_{\text{S}}$. The other inputs to $\text{BDT}_{\text{S}}$ are $\Delta E$, the distribution of which peaks at $0 \gev$ for signal events, and vertex information for the dielectron system, as true $\epen$ pairs in a signal event will originate from the same point. The agreement between data and MC for the inputs and the output of $\text{BDT}_{\qqbar}$ and $\text{BDT}_{\text{S}}$ is verified using the control modes $\B \to \Kstar c\bar{c}(\to \epen)$ and $B \to \Kstar \gamma (\to \epen)$ and the $\mbc < 5.27 \gevcc$ sideband of $\BtoKstee$ candidates. Here, $c\bar{c}$ candidates are reconstructed from dielectron candidates with an invariant mass within a $1\sigma$ window of the mass of the $\jpsi$ and $\psitwos$ resonances~\cite{cite-PDG}, while $\gamma$ candidates, which are reconstructed from dielectron candidates created in photon conversions in the detector material, are required to have $q^2 < 0.1 \gevccsq$ and have a dielectron vertex radius in the $x$-$y$ plane of at least $0.52 \cm$. The output of $\text{BDT}_{\qqbar}$ and $\text{BDT}_{\text{S}}$ for signal-like $\B \to \Kstar c\bar{c}$ events and the background-like $\mbc$ sideband events are shown in Fig.~\ref{fig:paper_BDTs}.

Deviations between data and simulations are found in the side-band sample, and a study for any potential bias is conducted in Sec.~\ref{sec:sys_uncert}.

\subsection{Selection criteria optimization}
To reduce $\B \to \Kstar \gamma$ background events that peak in the signal $\mbc$ region, a $q^2$ lower threshold and a dielectron vertex radius upper threshold are simultaneously determined by maximizing the FOM from Eq.~\ref{eq:fom}. In this case, the background is taken to be the remaining simulated $\B \to \Kstar \gamma$ events. The $q^2$ lower threshold is determined to be $0.0008 \gevccsq$, coincidentally the same as LHCb~\cite{LHCb:2020dof}, and the radius upper threshold is $0.52 \cm$.

After applying these criteria, the lower threshold for $\text{BDT}_{\text{S}}$ is determined to be $-0.79$ by maximizing the FOM against all remaining background. Taking all selection criteria into account, for the $\Kstarz \to K^+ \pi^-$ ($\Kstarp \to \KS \pi^+$) channel, approximately $18$ ($5$) simulated signal events and $16$ ($8$) background events are expected to remain in the $\mbc > 5.27 \gevcc$ region, corresponding to a signal reconstruction efficiency of $10.0\%$ ($5.8\%$). This is a $90\%$ increase in signal yield with similar remaining background when compared to using the same selection criteria but replacing the electron identification criteria with the often used binary LHR threshold of $0.9$. In this $\mbc$ region, $80\%$ of simulated background events have both electrons correctly identified. Continuum events are the largest background ($24\%$ of the remaining sample, or 11 events) in simulations for $\mbc > 5.27 \gevcc$, followed by lepton candidates that originated from different particles ($18\%$ or 9 events). Peaking background from $B\to K^* \gamma$ events contribute to about $2\%$, or 1 event, of the remaining simulated sample that will be used to develop the fitting procedure. 

\section{Analysis method}
A two-stage binned log-likelihood fit is performed, first for $\mbc$, and then simultaneously for projections of $\ctl$, $\ctk$ and $\phi$ to measure the free parameters in the differential decay rate given in Eq.~\ref{eq:diff_reduced}. 

\subsection{Fit to $\mbc$}
A one-dimensional fit to the $\mbc$ distribution is performed over the range $5.23 < \mbc < 5.29 \gevcc$ to determine the relative contributions of signal events, $f_{\rm S}$, peaking background events, $f_{\rm P}$, and all other background, $f_{\rm B}$. For signal and peaking background events, separate Gaussian PDFs are defined, where the means and widths are determined with simulations. For all other background, an ARGUS function~\cite{ALBRECHT1990278} is used. In the $\mbc$ fit, $f_{\rm P}$ is fixed to its expected value from simulations, $0.02$, and all PDF shape parameters are fixed to the values determined in fits to each category separately. The value of $f_{\rm B}$ remains floating, with an expected value of $0.85$, and is used after the $\mbc$ fit to calculate the relative contributions of each category in the $\mbc > 5.27 \gevcc$ signal region.

\subsection{Fit to the angular distributions}
The PDF for the projections of $\ctl$, $\ctk$ and $\phi$ for signal events in the $\mbc > 5.27 \gevcc$ region is defined as a product of the differential decay rate from Eq.~\ref{eq:diff_reduced} and factorized acceptance functions for each angle~\cite{LHCb:2015ycz}. The efficiency of signal events as a function of $\phi$ is expected to be uniform, whereas the acceptance functions for $\ctl$ and $\ctk$ are defined as fourth-order Legendre polynomials. The coefficients are determined by generating signal events with uniform angular distributions and then fitting to the angular projections of events that remain after applying all selection criteria. The acceptance function for $\ctl$ is set to be symmetric. 

Histogram PDFs are defined using simulations of the peaking background and all other background in the $\mbc > 5.27 \gevcc$ region for each angular distribution. Cross-checks are performed for this choice of PDF definition, finding that results in MC are in agreement with fits using histogram PDFs defined using the $\mbc < 5.27 \gevcc$ sideband region in data.

The angular PDFs are then summed using the coefficients that were determined from the $\mbc$ fit. This combined PDF is then fit to the angular distributions, where $\ATt$ and $\ATI$ are floated, and $F_L$ and $\ATR$ are fixed to their value expected in SM, as determined by the flavio package~\cite{Straub:2018kue}. The SM values, while consistent with the LHCb result~\cite{LHCb:2020dof}, are used to maintain independence from previous measurements. The decision to fix these parameters was due to the limited statistical power of the dataset, and were chosen because although they may differentiate between new physics scenarios, their impact on $\csev^{(\prime)}$ constraints is minimal~\cite{Becirevic:2011bp,Kruger:2005ep}.

A pull test and a linearity test are performed to investigate potential bias from the $\mbc$ and angular fitting procedures. Pseudo-data sets are generated at the expected Belle luminosity using $f_{\rm B} = 0.85$, while for $\ATt$ and $\ATI$, their SM values are used in the pull test, or over the range $[-1,1]$ in the linearity test. The linearity test has a gradient consistent with unity and an intercept of zero, and the pull test finds a mean consistent with zero and a width consistent with one.  

\section{Systematic uncertainties}
\label{sec:sys_uncert}
Various sources of potential systematic error are investigated, with uncertainties quantified where effects are not deemed negligible. They are described in turn below, and summarized in Table~\ref{tab:sys_summary}.

Systematic uncertainties are assigned for the efficiency and mis-identification of $e^\pm$ candidates and charged long-lived hadrons. We propagate uncertainties on corrections for particle identification performance in data based on measurements of $\Dstar$ decays for $\mathcal{L}_{\pi/K}$~\cite{NAKANO2002402}, while the analysis detailed in Sec.~\ref{sec:paper_LID} is used to measure the LHR and $\text{BDT}_{e}$ corrections used to propagate the uncertainty for electron identification.

Potential bias from $\text{BDT}_{\text{S}}$ is investigated by measuring the ratio of the yield of $\B \to \Kstar c\bar{c}$ from fits to data and simulations as a function of the $\text{BDT}_{\text{S}}$ lower threshold. The ratio is found to be consistent in the region in which the threshold is applied to the $\BtoKstee$ analysis and therefore no systematic uncertainty is assigned for $f_{\rm B}$, and effects on $\ATt$ and $\ATI$ are absorbed into the uncertainties for fit parameters, described below.

Although we find a fit bias to be consistent with zero within uncertainties, we assign the Gaussian mean value of the pull test multiplied by the expected statistical uncertainty as a systematic uncertainty. Potential biases that are folded into the detector response are also investigated by using the BTOSLLNP event generator to generate signal events across a spectrum of $\csev^{(\prime)}$ magnitudes and complex arguments followed by performing a pull and linearity test.
The values for the four free parameters in Eq.~\ref{eq:diff_reduced} are calculated for each of these $\csev^{(\prime)}$ values using flavio~\cite{Straub:2018kue} and a covariance matrix is determined. New pseudo-data sets are then created using randomly generated values of the four parameters while taking correlations into account~\cite{Kaiser1962}, followed by a fit with $F_L$ and $\ATR$ fixed to their SM values. The width of the distributions of the $\ATI$ and $\ATt$ fit results quantify the uncertainty for fixing $F_L$ and $\ATR$ to their SM value. 

The fixed values for the $\mbc$ and acceptance PDF fit parameters are fluctuated according to their $1 \sigma$ uncertainty determined from the individual MC fits. Pseudo-data sets of size $10^4$ are then generated and the fitting procedure is applied to determine an associated systematic uncertainty using parameter distribution widths. Any bias from factorizing the acceptance function and using a flat angular distribution is measured using a method similar to Ref.~\cite{Borsato:2015zuf}. The dataset is re-weighted with the reciprocal of the acceptance function of $\ctl$ to recover the flat angular distribution in $\ctl$, it is then re-weighted with the reciprocal of $\ctk$. A Legendre polynomial is fit to this distribution resulting in coefficients consistent with zero. The same effect is seen in events generated using BTOSLLBALL and therefore no systematic uncertainty is applied for any bias from factorization.

For $f_{\rm P}$, the ratio of the yield of $B \to \Kstar \gamma (\to \epen)$ events between data and simulations is measured and found to be $0.96 \pm 0.07$. Pseudo-data sets are then generated by correcting the value of $f_{\rm P}$ according to the ratio and fluctuating it within a $1\sigma$ uncertainty.

Beam energy and magnetic field mismodeling is investigated by fitting $\B \to \Kstar c\bar{c}$ in data with the mean of the signal Gaussian and an overall width factor, $w$ allowed to remain floating. The results, $\mu = 5.27949 \pm 0.00003$ \gevcc and $w = 0.98 \pm 0.01$, are used to generate pseudo-data sets, for which the spread of parameter fit results is negligible and therefore no systematic uncertainty is assigned.

The uncertainty due to the limited number of simulated events is assigned by propagating the statistical error for each of the different event types, which were generated with different integrated luminosities between $5$ and $10$ times the size of the Belle data set.

\begin{table}[htb]
  \centering
  \caption{A summary of all the systematic uncertainties in the $\BtoKstee$ angular analysis.}
\begin{tabular}{lrrr}

           Uncertainty &  $\ATt$ &  $\ATI$ &  $f_{\rm B}$ \\
\hline
Electron ID efficiency &   0.004 &   0.004 &        0.001 \\
Electron mis-ID &   0.001 &   0.002 &        0.001 \\
Hadron ID efficiency &   0.002 &   0.001 &        0.000 \\
Hadron mis-ID &   0.003 &   0.003 &        0.001 \\
Fit bias &   0.014 &   0.016 &        0.002 \\
Signal generator &   0.039 &   0.059 &        0.000 \\
Fixed $F_L$ and $\ATR$ &   0.009 &   0.033 &        0.000 \\
Fixed $\mbc$ params. &   0.014 &   0.021 &        0.001 \\
Fixed acceptance params. &   0.022 &   0.015 &        0.000 \\
Fixed $f_{\rm P}$ &   0.008 &   0.012 &        0.000 \\
MC statistics &   0.097 &   0.095 &        0.011 \\   
\hline
Total & 0.109 & 0.121 & 0.011 
\end{tabular}
\label{tab:sys_summary}
\end{table}

\section{Results}
\begin{figure*}
    \centering
    \includegraphics[width=0.96\linewidth]{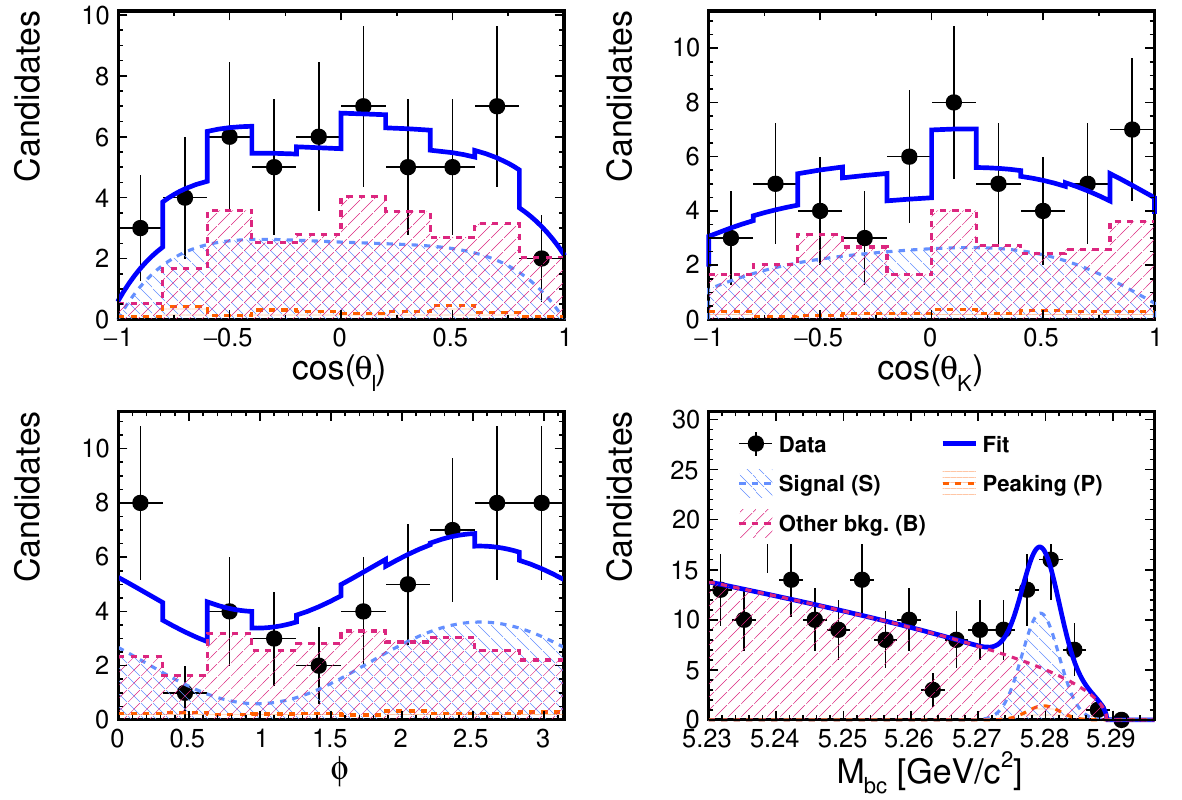}
    \caption{Projections of $\cos{\thetal}$, $\cos{\thetaK}$, $\phi$, and $\mbc$ are shown, determined from the fits to the full Belle data set, noting that only $\ATt$ and $\ATI$ are floated in the angular PDF. The angular distributions are plotted for the $\mbc > 5.27\gevcc$ signal region.}
    \label{fig:data_fit}
\end{figure*}

The fitting procedure finds a yield of $21 \pm 6$ signal events and $29 \pm 5$ background events for $\mbc > 5.27 \gevcc$, corresponding to a value of $f_{\rm B} = 0.86 \pm 0.04 \pm 0.01$, which is in agreement with simulations. Performing a profile likelihood test on this result, with the systematic uncertainties included, returns a significance of $3.3\sigma$ against the background-only hypothesis. The fits to $\mbc$ and the projections of the three angular observables are shown in Fig.~\ref{fig:data_fit}.

The results for the floated transversality amplitudes in the $0.0008 < q^2 < 1.12 \gevccsq$ range are
\begin{equation}
    \begin{split}
        \ATt &= 0.52 \pm 0.53 \pm 0.11, \\
        \ATI &= -1.27 \pm 0.52 \pm 0.12,
    \end{split}
\end{equation}
where the first uncertainty is statistical and the second is systematic. The correlation between these two values is found to be $0.10$.

\begin{figure*}
    \centering
    \includegraphics[width=0.7\linewidth]{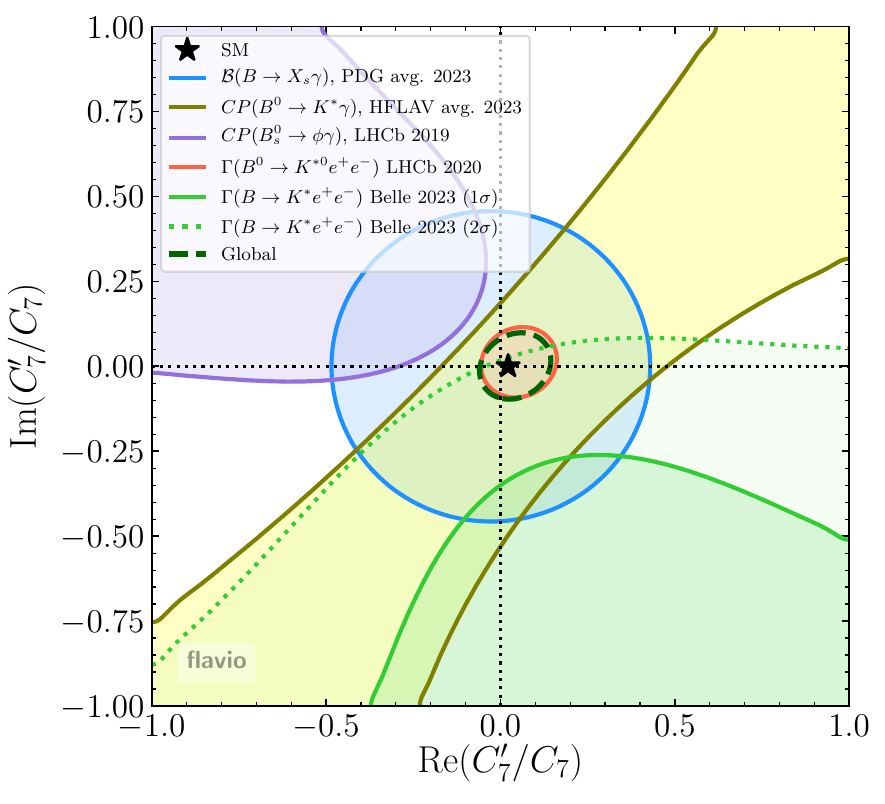}
    \caption{Contours for confidence limits (at $1\sigma$ unless stated otherwise) on $\csev^\prime / \csev$ are shown for the angular analyses in this paper, the LHCb 2020 study~\cite{LHCb:2020dof}, and for several other branching fraction and $CP$-violation parameter measurements~\cite{LHCb:2019vks,PhysRevD.107.052008,cite-PDG}. These constraints are determined in flavio~\cite{Straub:2018kue} with $\csev = -0.2915$.}
    \label{fig:flav_constraints_data}
\end{figure*}

Constraints on the ratio of $\csev^\prime / \csev$ are shown in Fig.~\ref{fig:flav_constraints_data}, with the left-handed Wilson coefficient fixed to its SM value, $\csev = -0.2915$~\cite{Blake:2016olu}. The figure includes constraints from the measurement of $CP$-violation parameters in $B^0_s \to \phi \gamma$ decays at LHCb~\cite{LHCb:2019vks}, $S= 0.43 \pm 0.30 \pm 0.11$ and $\mathcal{A}^\Delta = -0.67 ~^{+0.37}_{-0.41} \pm 0.17$, and $B\to K^* \gamma$ or $B\to K^0 \pi^0 \gamma$ decays averaged by HFLAV~\cite{PhysRevD.107.052008}, $A_{CP} = -0.006 \pm 0.011$ and $S = -0.16 \pm 0.22$. Also shown are the inclusive branching fraction $\mathcal{B}(B\to X_s \gamma) = (3.49 \pm 0.19) \times 10^{-4}$~\cite{cite-PDG}, and the $\BztoKstee$ angular analysis from LHCb~\cite{LHCb:2020dof}, with $\ATt = 0.11 \pm 0.10 \pm 0.02$ and $\ATI = 0.02 \pm 0.10 \pm 0.01$. The results with and without the LHCb measurement are combined into a global $1\sigma$ constraint. 

The results from this analysis alone constrain the value of $\csev^\prime$ to be in agreement with the SM within a $2\sigma$ confidence interval, and the combined results remain consistent with the SM, independent of the LHCb measurement.

\section{Conclusion}
An angular analysis of $\BtoKstee$ decays for a dielectron invariant mass squared range of $0.0008-1.12 \gevccsq$ is performed using techniques in electron identification that are new to Belle. This new electron identification uses machine learning and reduces electron mis-identification rates by up to a factor $5$, expanding the capabilities of rare $B$ decay searches at the experiment. The angular analysis finds $\ATt = 0.52 \pm 0.53 \pm 0.11$ and $\ATI = -1.27 \pm 0.52 \pm 0.12$. This constrains non-SM right-handed contributions with the finding that $\csev^\prime$ is in agreement with the SM expectation within a $2\sigma$ confidence interval. Scaling these results to the design luminosity of Belle II will mean future measurements will have statistical sensitivity competitive with that of LHCb.

\section*{Acknowledgments}
This work, based on data collected using the Belle detector, which was
operated until June 2010, was supported by 
the Ministry of Education, Culture, Sports, Science, and
Technology (MEXT) of Japan, the Japan Society for the 
Promotion of Science (JSPS), and the Tau-Lepton Physics 
Research Center of Nagoya University; 
the Australian Research Council including grants
DP210101900, % Urquijo
DP210102831, % Sevior
DE220100462, % Hsu
LE210100098, % Infrastructure
LE230100085; % Infrastructure
Austrian Federal Ministry of Education, Science and Research (FWF) and
FWF Austrian Science Fund No.~P~31361-N36;
National Key R\&D Program of China under Contract No.~2022YFA1601903,
National Natural Science Foundation of China and research grants
No.~11575017,
No.~11761141009, 
No.~11705209, 
No.~11975076, 
No.~12135005, 
No.~12150004, 
No.~12161141008, 
and
No.~12175041, 
and Shandong Provincial Natural Science Foundation Project ZR2022JQ02;
the Ministry of Education, Youth and Sports of the Czech
Republic under Contract No.~LTT17020;
the Czech Science Foundation Grant No. 22-18469S;
Horizon 2020 ERC Advanced Grant No.~884719 and ERC Starting Grant No.~947006 ``InterLeptons'' (European Union);
the Carl Zeiss Foundation, the Deutsche Forschungsgemeinschaft, the
Excellence Cluster Universe, and the VolkswagenStiftung;
the Department of Atomic Energy (Project Identification No. RTI 4002) and the Department of Science and Technology of India; 
the Istituto Nazionale di Fisica Nucleare of Italy; 
National Research Foundation (NRF) of Korea Grant
Nos.~2016R1\-D1A1B\-02012900, 2018R1\-A2B\-3003643,
2018R1\-A6A1A\-06024970, RS\-2022\-00197659,
2019R1\-I1A3A\-01058933, 2021R1\-A6A1A\-03043957,
2021R1\-F1A\-1060423, 2021R1\-F1A\-1064008, 2022R1\-A2C\-1003993;
Radiation Science Research Institute, Foreign Large-size Research Facility Application Supporting project, the Global Science Experimental Data Hub Center of the Korea Institute of Science and Technology Information and KREONET/GLORIAD;
the Polish Ministry of Science and Higher Education and 
the National Science Center;
the Ministry of Science and Higher Education of the Russian Federation, Agreement 14.W03.31.0026, % from 15.02.2018
and the HSE University Basic Research Program, Moscow; % from 15.04.2021
University of Tabuk research grants
S-1440-0321, S-0256-1438, and S-0280-1439 (Saudi Arabia);
the Slovenian Research Agency Grant Nos. J1-9124 and P1-0135;
Ikerbasque, Basque Foundation for Science, Spain;
the Swiss National Science Foundation; 
the Ministry of Education and the Ministry of Science and Technology of Taiwan;
and the United States Department of Energy and the National Science Foundation.
These acknowledgements are not to be interpreted as an endorsement of any
statement made by any of our institutes, funding agencies, governments, or
their representatives.
We thank the KEKB group for the excellent operation of the
accelerator; the KEK cryogenics group for the efficient
operation of the solenoid; and the KEK computer group and the Pacific Northwest National
Laboratory (PNNL) Environmental Molecular Sciences Laboratory (EMSL)
computing group for strong computing support; and the National
Institute of Informatics, and Science Information NETwork 6 (SINET6) for
valuable network support.

\bibliography{references}

\end{document}

%% file: pub665-orcid.tex
 %% Paper:    B to K* e e
%%% Journal:  Physical Review D
%%% Contacts: D. Ferlewicz, P. Urquijo
%%% Non-responding authors or those who said NO are commented out.
%%% ====================================================================
%%% Click the RELOAD button on your web browser to see the updated file.
%%% ====================================================================
%%% Use \input{pub665-orcid} to insert this material into your latex file.
\noaffiliation
\author{D.~Ferlewicz\,\orcidlink{0000-0002-4374-1234}} % 2073
  \author{P.~Urquijo\,\orcidlink{0000-0002-0887-7953}} % 2302
  \author{I.~Adachi\,\orcidlink{0000-0003-2287-0173}} % 2590
  \author{K.~Adamczyk\,\orcidlink{0000-0001-6208-0876}} % 2239
% \author{J.~K.~Ahn\,\orcidlink{0000-0002-5795-2243}} % 7423
  \author{H.~Aihara\,\orcidlink{0000-0002-1907-5964}} % 2223
% \author{S.~Al~Said\,\orcidlink{0000-0002-4895-3869}} % 6823
  \author{D.~M.~Asner\,\orcidlink{0000-0002-1586-5790}} % 4684
  \author{H.~Atmacan\,\orcidlink{0000-0003-2435-501X}} % 2538
% \author{V.~Aulchenko\,\orcidlink{0000-0002-5394-4406}} % 8183
% \author{T.~Aushev\,\orcidlink{0000-0002-6347-7055}} % 3747
  \author{R.~Ayad\,\orcidlink{0000-0003-3466-9290}} % 3766
% \author{T.~Aziz\,\orcidlink{-}} % 3523
  \author{V.~Babu\,\orcidlink{0000-0003-0419-6912}} % 5623
% \author{S.~Bahinipati\,\orcidlink{0000-0002-3744-5332}} % 2332
% \author{A.~M.~Bakich\,\orcidlink{0000-0001-8315-4854}} % 2115
% \author{Y.~Ban\,\orcidlink{-}} % 3503
  \author{Sw.~Banerjee\,\orcidlink{0000-0001-8852-2409}} % 8603
% \author{E.~Barberio\,\orcidlink{-}} % -229
% \author{M.~Barrett\,\orcidlink{0000-0002-2095-603X}} % 2180
% \author{M.~Bauer\,\orcidlink{0000-0002-0953-7387}} % 9863
% ...
% 2387
% \author{V.~Zhulanov\,\orcidlink{0000-0002-0306-9199}} % 4983
\collaboration{The Belle Collaboration}